\documentclass[conference,twocolumn]{IEEEtran}
\IEEEoverridecommandlockouts
\usepackage{amsfonts}
\usepackage{amssymb}
\usepackage{cite}
\usepackage[cmex10]{amsmath}
\usepackage{float}
\usepackage{color}
\usepackage{stfloats,fancyhdr}
\usepackage{amsmath}
\usepackage{amsthm}
\usepackage{algorithm}
\usepackage{algorithmic}
\usepackage{multirow}
\usepackage{changepage}
\usepackage[normalem]{ulem}
\usepackage{booktabs}

\usepackage{url}

\IEEEoverridecommandlockouts

\ifCLASSINFOpdf
  \usepackage[pdftex]{graphicx}
  \DeclareGraphicsExtensions{.pdf,.jpeg,.png}
\else
  \usepackage[dvips]{graphicx}
  \DeclareGraphicsExtensions{.pdf}
\fi

\usepackage{subfigure}

\usepackage{fancybox,dashbox}

\begin{document}
%
\title{{\huge Fresh-Fi: Enhancing Information Freshness in Commodity WiFi Systems via Customizing Lower Layers 
}}


\author{\IEEEauthorblockN{Zixiao Han, Qian Wang and He (Henry) Chen}
   \thanks{This work is supported in part by the Innovation and Technology Fund (ITF) under Project ITS/204/20 and the CUHK direct grant for research under Project 4055166. 
}
 \IEEEauthorblockA{Department of Information Engineering, The Chinese University of Hong Kong, Hong Kong SAR, China} 
 Email: \{hz022, qwang, he.chen\}@ie.cuhk.edu.hk}



\maketitle
\vspace{-1em}

\begin{abstract}
Enhancing information freshness in wireless networks has gained significant attention in recent years. To optimize or analyze information freshness, which is often characterized by the age of information (AoI) metric, extensive theoretical studies have been conducted on various wireless networks. Early research has demonstrated the significance of last-come-first-served (LCFS) packet scheduling and controlled status sampling (i.e., packet generation) in improving information freshness. These mechanisms have been widely adopted in subsequent studies. However, the effective implementation of these mechanisms in commercial off-the-shelf (COTS) wireless devices has not been thoroughly investigated, which could limit the practical application of information freshness-oriented protocols in real-world systems. Our work aims to address the gap by exploring the effective implementation of the information freshness-oriented mechanisms mentioned above in COTS WiFi devices that use the Linux operating system. Our attempts reveal that implementing these mechanisms in COTS systems is not a straightforward task. Specifically, we found that the physical layer queue of WiFi devices operates on a first-come-first-served (FCFS) basis, and the packet generation process cannot be precisely controlled by default. To overcome these challenges, we develop Fresh-Fi, an information freshness-oriented protocol stack that involves careful customization to the lower layers of the Linux networking protocol stack. Fresh-Fi mainly incorporates a mac80211 subsystem-based LCFS queue and a real-time kernel-based cross-layer tunnel between the mac80211 subsystem and the application layer for triggered packet generation. Our experiments show that implementing Fresh-Fi can significantly improve AoI performance. Specifically, we observed that Fresh-Fi improved AoI performance by over 13 times when compared to a baseline design that relies on an LCFS queue implemented in the application layer of the standard Linux.

\textit{Index Terms-}Information freshness, age of information, commercial off-the-shelf WiFi devices, Linux kernel, mac80211 subsystem.

\end{abstract}

\IEEEpeerreviewmaketitle

\section{Introduction}

The emergence of the Internet of Things (IoT) has offered a promising paradigm in supporting massive connections among devices in various networks, including but not limited to the IEEE 802.11 standard (WiFi) \cite{Agriculture,drone}. Generally, IoT involves the networked interconnection of objects/things/sensors/devices to achieve performance increment \cite{xia2012internet, perera2015emerging}. In time-sensitive IoT applications, the freshness of the information between the source node and destination node has become a primary concern. To solve this concern, the age of information (AoI) is proposed as a metric to characterize information freshness. It is defined as the time elapsed since the generation time of the latest successfully received status update at the destination node\cite{Kaul2011mini}.

To optimize AoI performance, it is important for the source node to deliver its status updates to the destination node as timely as possible \cite{how_often}. This cannot be achieved by simply minimizing the transmission delay for each status update, since AoI continues to rise between the arrival of two consecutive status updates. Meanwhile, timely status updating is not the same as generating the status update as frequently as possible. This is because in real-world wireless networks, the source node stores the newest status update in a transmission queue awaiting transmission until all previous packets have been successfully delivered. The backlogging time suffered by the stream of status updates increases with the accumulation of packets in the queue. 

To avoid the long waiting time in the queue, prior research suggests two effective mechanisms. First, as an alternative to the first-come-first-served (FCFS) policy \cite{how_often}, the last-come-first-served (LCFS) policy can be applied to manage the transmission queue for enhancing AoI performance. It improves AoI performance by discarding the stale status updates stored in the transmission queue, ensuring that the latest status update can be transmitted first \cite{LCFS_BETTER, Status_update_LCFS}. Second, integrating a proactive status sampling policy that controls the generation of status update packets is another option \cite{ Age_penalty_algorithm}. This policy is based on the assumption of a zero-delay acknowledgment (ACK) sent from the destination node, which informs the status sampling
policy about the transmission completion of the previous status update \cite{What_is_just_in_time_updating,zero_wait_not_optimum}. It ensures that the source node schedules the next status update generation only after the previous one has been delivered. The simplest yet most widely used policy is the zero-wait policy proposed in \cite{What_is_just_in_time_updating} wherein the source node generates and sends a status update immediately after ACK reception. As a result, the backlogging time for status updates is significantly reduced by keeping the number of packets stored in the transmission queue no greater than one and controlling the generation of the next status update once one is delivered. 

While the effectiveness of the above two mechanisms for improving AoI performance has been well-proven in the theoretical literature, their implementation and evaluation in commercial off-the-shelf (COTS) communication systems have been limited. A few studies have attempted to improve AoI performance by modifying the application layer or the transport layer on COTS WiFi devices \cite{WiFresh,Wiswarm,Transport_layer_AoI_protocol}. Nevertheless, their experimental results are far from optimal due to the untouched lower layers in the protocol stack, which act as a bottleneck for further improving the AoI performance. This has been confirmed by our experiments presented in Sec. \ref{Experiment results}. Another line of research \cite{Zixiao,Push_and_pull,Semantics,Germany_NCS} implemented the information freshness-oriented lower layer protocols on software-defined radio (SDR) platforms, which avoids the complexity of customizing protocol layers on actual hardware. While these studies offer valuable insights for optimizing information freshness in practical networks, their designs are not directly applicable in practice due to the high cost of SDR platforms and the resultant limited scalability in large-scale systems. To our best knowledge, there is a lack of comprehensive research on customizing lower protocol stack layers in COTS WiFi devices, which are typically integrated into the operating system kernel, to incorporate mechanisms aimed at improving information freshness. We note that making modifications to the kernel networking stack is not a trivial task because of its integrity and complexity.

To address this gap, we introduce Fresh-Fi, a customized protocol stack designed to enhance information freshness in COTS WiFi systems that use the Linux operating system. We chose to focus on optimizing the AoI in Linux-based COTS WiFi systems due to their widespread adoption in time-sensitive industrial sectors and the open-source nature of the Linux kernel. {Fresh-Fi consists of two main components: 1) an LCFS queue that prioritizes the status update transmissions, and 2) a cross-layer tunnel that is designed to relay the transmission completion notification from the mac80211 subsystem to the status sampling policy in the application layer for scheduling the next status generation. To evaluate the efficacy of Fresh-Fi, we compare the AoI performance of Fresh-Fi against two baseline designs. The experiment results indicate that the implementation of Fresh-Fi improves AoI performance by over 20-fold when compared to the simple design that uses the standard user datagram protocol (UDP) for delivering status updates. Furthermore, implementing Fresh-Fi improves AoI performance by over 13 times when compared to the design proposed in \cite{WiFresh}, which relies on an application-layer LCFS queue and a destination-based polling mechanism.} We remark that since Fresh-Fi does not touch the driver layer, it is compatible with any WiFi network interface cards that are supported by the Linux mac80211 subsystem. In addition, while Fresh-Fi is primarily designed to optimize the AoI of COTS WiFi systems by modifying the Linux protocol stack, its design principles can be extended to other COTS wireless devices that use different operating systems.

\section{Preliminaries and Motivations}\label{Preliminaries}
\subsection{A Primer on AoI}
\begin{figure}
    \centering
    \includegraphics[width=0.4\textwidth]{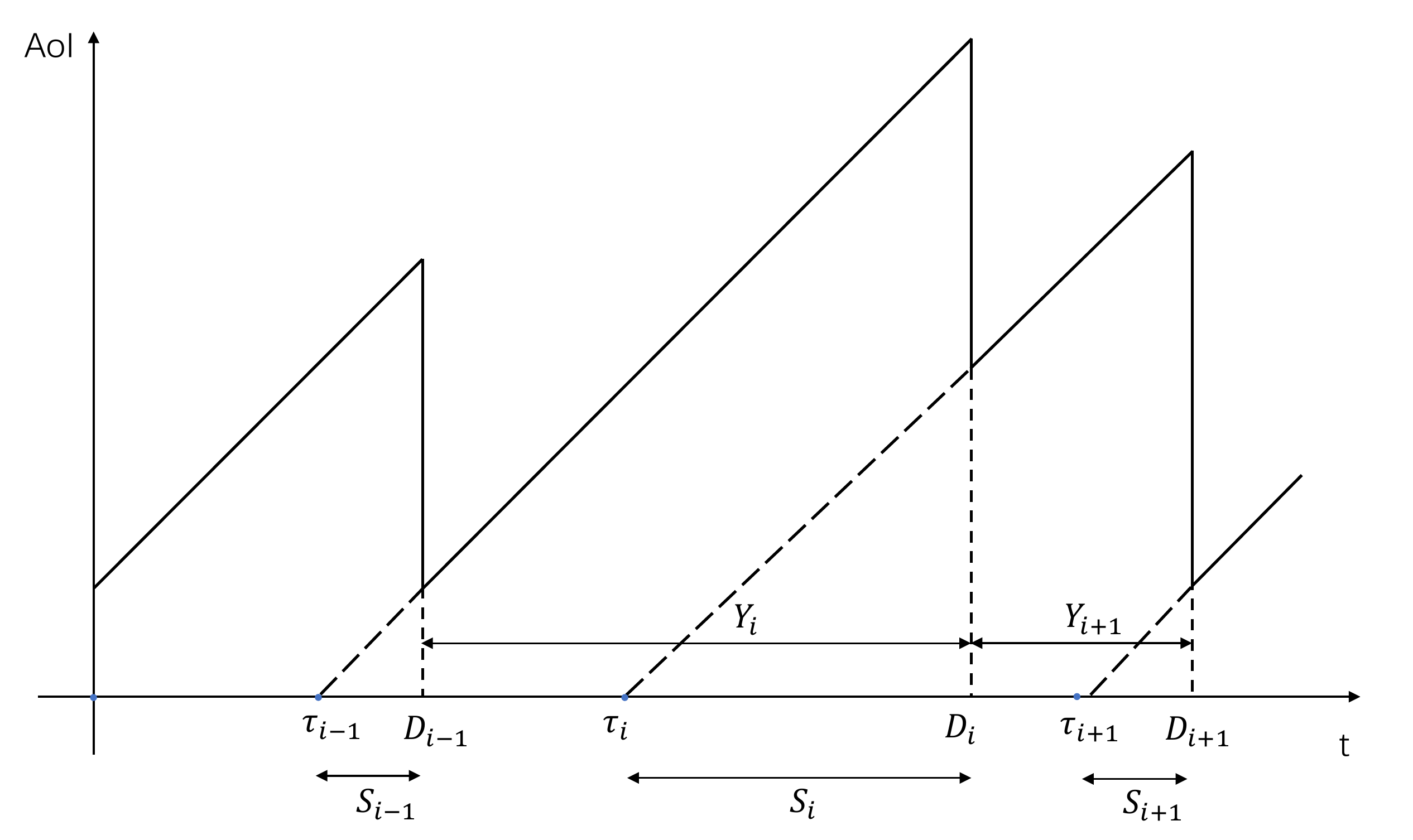}
    \caption{An AoI evolution for a source node sending status updates to the destination node.}
   \label{fig: Ins_AoI}
   \vspace{-1em}
\end{figure}
The AoI is defined as the time elapsed since the generation time of the most recently received status update \cite{how_often}. At the time $t$, the destination node receives the $i${th} status update generated at time $\tau_i(t)$. Then the AoI at the destination node is calculated as $A(t):=t-\tau_i(t)$. Let $D_i$ denote the time at which the $i${th} status update is received. The service time of the $i${th} status update packet is defined as $S_i = D_i - \tau_i$. The interval between two consecutive receptions is denoted by $Y_i$, and can be calculated as $Y_i = D_i - D_{i-1}$. According to \cite{Henry_AoI_calculation}, the average AoI can be expressed as $\overline{A} = \frac{1}{\mathbb{E}[Y_i]}(\mathbb{E}[S_{i-1}Y_i] + \frac{\mathbb{E}[Y_i^2]-\mathbb{E}[Y_i]}{2})$. Fig. \ref{fig: Ins_AoI} presents an example of the instantaneous AoI evolution. From this figure, we can see that the AoI only decreases at the moments when the destination node receives a new status update. During the interval between two consecutive status update generations and the status update service time, the AoI keeps increasing. It reveals that the average AoI is affected by both the status update generation and transmission processes.


To optimize AoI performance, ideally, the source node should schedule the next status generation immediately after completing the transmission of the previous status update. Additionally, it is vital to reduce the service time of the latest delivered status update as much as possible. However, in practice, the default Linux networking protocol stack cannot meet these requirements. In the next subsection, we discuss the Linux networking protocol stack in detail and explain its limitations that have to be addressed to facilitate timely status updates from the source node to the destination node.

\subsection{A Primer on the Linux Networking Protocol Stack}\label{subsection_Kernl_stack}

\begin{figure}
    \centering
    \includegraphics[width=0.35\textwidth]{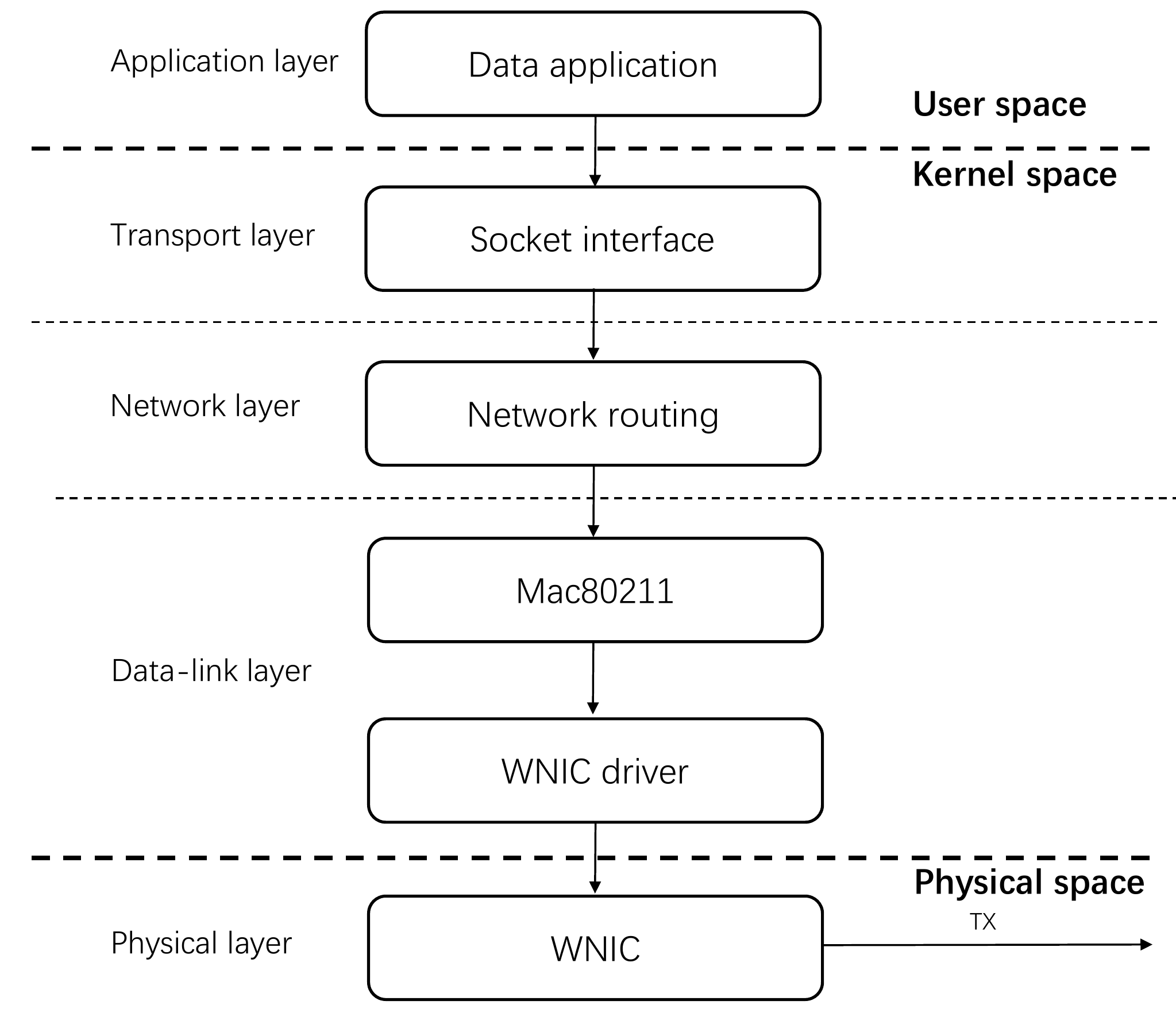}
    \caption{The diagram of the Linux networking protocol stack architecture.}
   \label{fig: Kernel stack}
   \vspace{-1em}
\end{figure}

As one of the most widely deployed wireless LAN protocols in the family of IEEE 802 Local Area Network standards \cite{IEEE80211}, IEEE 802.11 standard specifies the data link layer and the physical layer protocols that facilitate wireless communications. In order to comply with this standard, the Linux operating system has implemented the Linux networking protocol stack, which is a set of software components that enables the interactions among the user, kernel, and physical spaces. Fig. \ref{fig: Kernel stack} depicts the architecture of the Linux networking protocol stack \cite{Linux_wireless_stack}. To improve clarity, we describe how each component delivers a data packet in a top-down manner, to better understand their functionalities.


The user space data application generates a data packet and sends it to the kernel space socket interface subsystem via system socket calls. The socket interface subsystem executes the transport layer functionalities by offering the transmission control protocol (TCP) and UDP sockets for packet transmissions. Once the transport layer header is included in the packet, the network routing subsystem, which is designed according to the network layer protocols, determines the suitable WiFi network interface card (WNIC) to transmit this packet to its destination node. The packet is then received by the mac80211 subsystem, which is a generic framework implemented in the data-link layer that facilitates communication between the kernel and physical spaces\cite{What_is_mac80211}. It offers crucial functions for the WNIC driver to perform communication operations such as packet transmission, and registers the WNIC driver callback functions for making specific configurations on the physical layer processes. Eventually, the mac80211 subsystem passes the packet to the WNIC driver for physical layer transmission.

The FCFS queue mechanism with multiple priorities implemented in the WNIC has a significant impact on the order in which packets are transmitted \cite{Priority_Queue}. At first, packets are placed into different queues based on their priority levels. An FCFS queue with a higher priority will be processed by the WNIC earlier than FCFS queues with lower priorities to ensure its timeliness in packet transmissions. Additionally, the WNIC employs the carrier sense multiple access with collision avoidance (CSMA/CA), which applies a random delay before packet transmission, to prevent collisions between devices communicating on the same channel. At the destination node, the WNIC deploys a receive interruption mitigation mechanism \cite{Data_sheet} to prevent the central processing unit (CPU) from getting overwhelmed during processing the incoming traffic. This mechanism generates only one interruption signal to the CPU for processing multiple packets received in a pre-defined time period. This causes the user space packet receiver program to receive packets in bursts, which harms AoI performance. Although the majority of WNICs and their drivers are not open-source, the mac80211 subsystem in the Linux networking protocol stack allows customization of the physical layer transmission process by making specific configurations.

\begin{figure}
    \centering
    \includegraphics[width=0.4\textwidth]{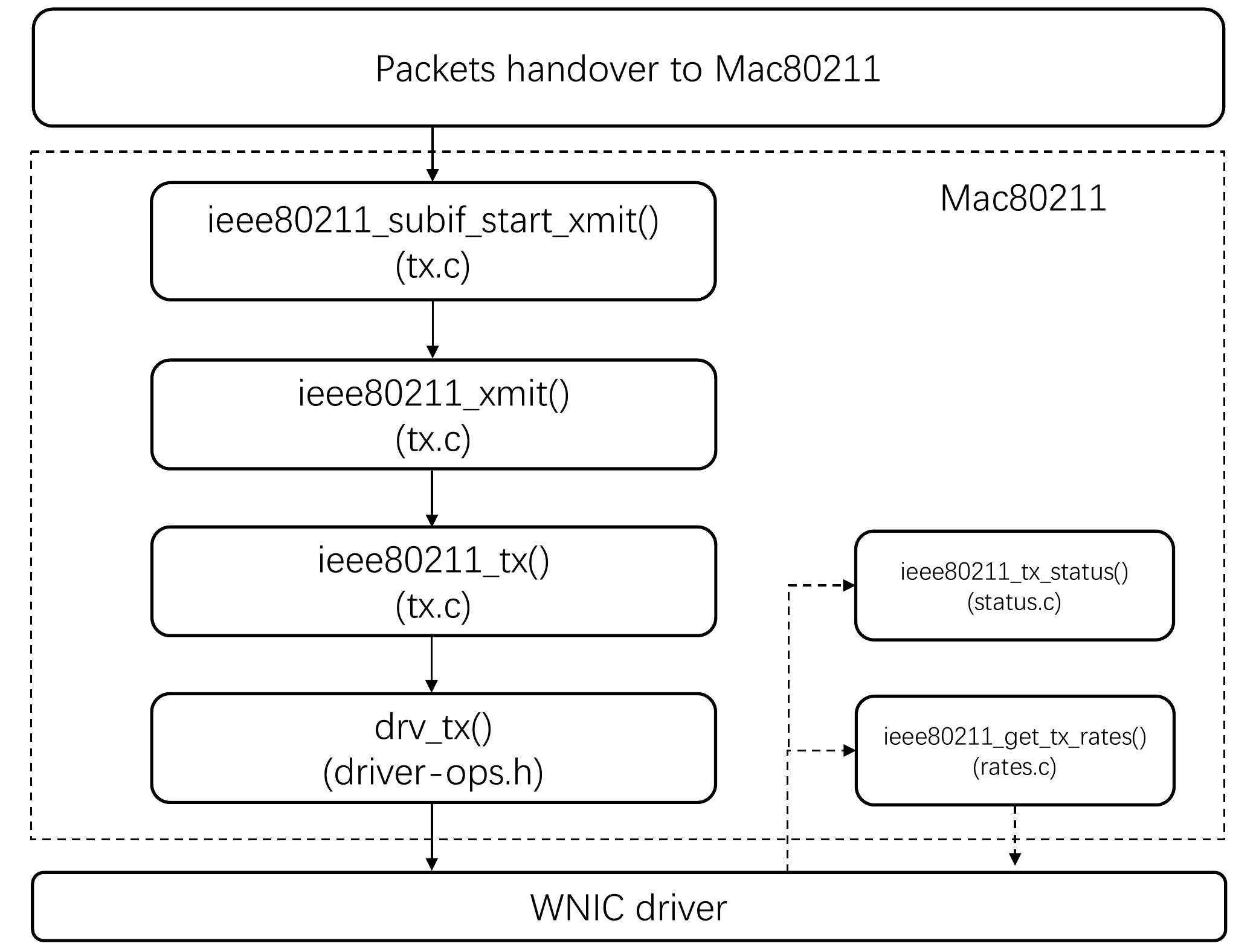}
    \caption{The packet transmission workflow in mac80211 subsystem.}
   \label{fig: Mac80211 details}
   \vspace{-1em}
\end{figure}


The inner structure of the mac80211 subsystem is presented by depicting its workflow in the packet transmission process, as illustrated in Fig. \ref{fig: Mac80211 details}. The \texttt{ieee80211\_subif\_start\_xmit()} function locates at the entrance of the mac80211 subsystem. It is responsible for receiving the packets handover from the upper layers. The \texttt{ieee80211\_xmit()} function then writes the data-link layer and the physical layer information to the headroom of the received packet. Next, the \texttt{ieee80211\_tx()} function passes the received packet to the FCFS queue belonging to the WNIC selected by the network routing subsystem. Finally, once the WNIC is ready for the next transmission, the mac80211 subsystem calls the {\texttt{drv\_tx()}} function to retrieve the first packet from the FCFS queue and forward it to the WNIC driver.

When a packet is received for delivery, the WNIC driver relies on several functions defined in the mac80211 subsystem to execute the packet transmission process. To enhance wireless transmission reliability, the WNIC driver uses the adaptive data rate adjustment algorithm by calling the \texttt{ieee80211\_get\_tx\_rates()} function. This algorithm generates a re-transmission table for each packet based on the current channel condition measured by the WNIC. This table provides a range of available data rates, and sets the maximum re-transmission times under each data rate. According to the re-transmission table, the WNIC repeatedly sends the same packet from the highest to lowest data rate until it receives a physical layer ACK from the destination node. If the WNIC has tried all the available data rates but still cannot receive the physical layer ACK, it tags this transmission as unsuccessful in the transmission status. Otherwise, this transmission is tagged as successful. Once the packet transmission completes, the WNIC calls the \texttt{ieee80211\_tx\_status()} function to generate an interrupt to report the transmission status to the mac80211 subsystem.

\subsection{Motivations to Design Fresh-Fi and to Implement it in the Mac80211 Subsystem}
Previous literature has suggested that implementing either the LCFS queue policy or the status sampling policy can considerably enhance AoI performance. Unfortunately, the Linux networking protocol stack does not support either of these policies by default. The LCFS queue policy requires that in the event of a transmission opportunity, only the most recently injected packet can be transmitted and the other stale packets are discarded. Nevertheless, the WNIC only possesses FCFS queues and it does not permit any packet drop during transmissions. The status sampling policy works by assuming that it will receive a zero-delay ACK to detect the previous status updating completion. However, this form of delay-free ACK transmission is unfeasible in real-world wireless networks. Specifically, the transport layer provides TCP and UDP for packet transmission. Upon receiving a packet, TCP controls the destination node to send a TCP ACK to the source node. However, the re-transmission and in-order delivery mechanisms of TCP cause longer transmission delays, which have a negative impact on AoI performance \cite{Transport_layer_AoI_protocol}. Compared with TCP, UDP achieves shorter transmission delay by ignoring transmission errors and continuing to transmit subsequent packets. Nevertheless, UDP does not offer an ACK for the status sampling policy to notify the completion of the previous status updating.

To bridge the gap between theoretical designs and practical implementations in COTS WiFi devices, in this paper, we propose Fresh-Fi, an information freshness-oriented protocol stack that customizes the Linux networking protocol stack. Fresh-Fi mainly comprises two components: a mac80211 subsystem-based LCFS queue that specifically serves the status update transmissions and a real-time kernel-based cross-layer tunnel between the mac80211 subsystem and the application layer for triggering status generations. Implementing Fresh-Fi in the mac80211 subsystem provides the following advantages:

 \begin{enumerate}
    \item \textbf{Good capability.} The mac80211 subsystem is designed to support various WNICs and their corresponding drivers sold on the market. By integrating the Fresh-Fi architecture into the mac80211 subsystem, Fresh-Fi can utilize this compatibility.
    \item \textbf{Re-transmission control.} The adaptive data rate adjustment algorithm is deployed in the mac80211 subsystem. By implementing here, Fresh-Fi can directly modify the source code of this algorithm to prohibit the WNIC from repeatedly delivering the stale status update to the destination node.

    \item \textbf{Access to physical layer configurations.} Although the WNIC and its driver are not open-source, they still register numerous callback functions and configurations to the mac80211 subsystem. As a result of utilizing these properties, Fresh-Fi can indirectly control the packet transmission process in the WNIC.

    \item \textbf{Access to transmission status.} The mac80211 subsystem is able to receive the transmission status from the WNIC, which indicates the success or failure of the previous packet transmission. Fresh-Fi can utilize this message to inform the status sampling policy about the completion of the previous status update transmission.
     
\end{enumerate}

\section{Mac80211 Subsystem-based LCFS Queue in Fresh-Fi}\label{mac80211 subsystem-based LCFS}

\begin{figure}
    \centering
    \includegraphics[width=0.45\textwidth]{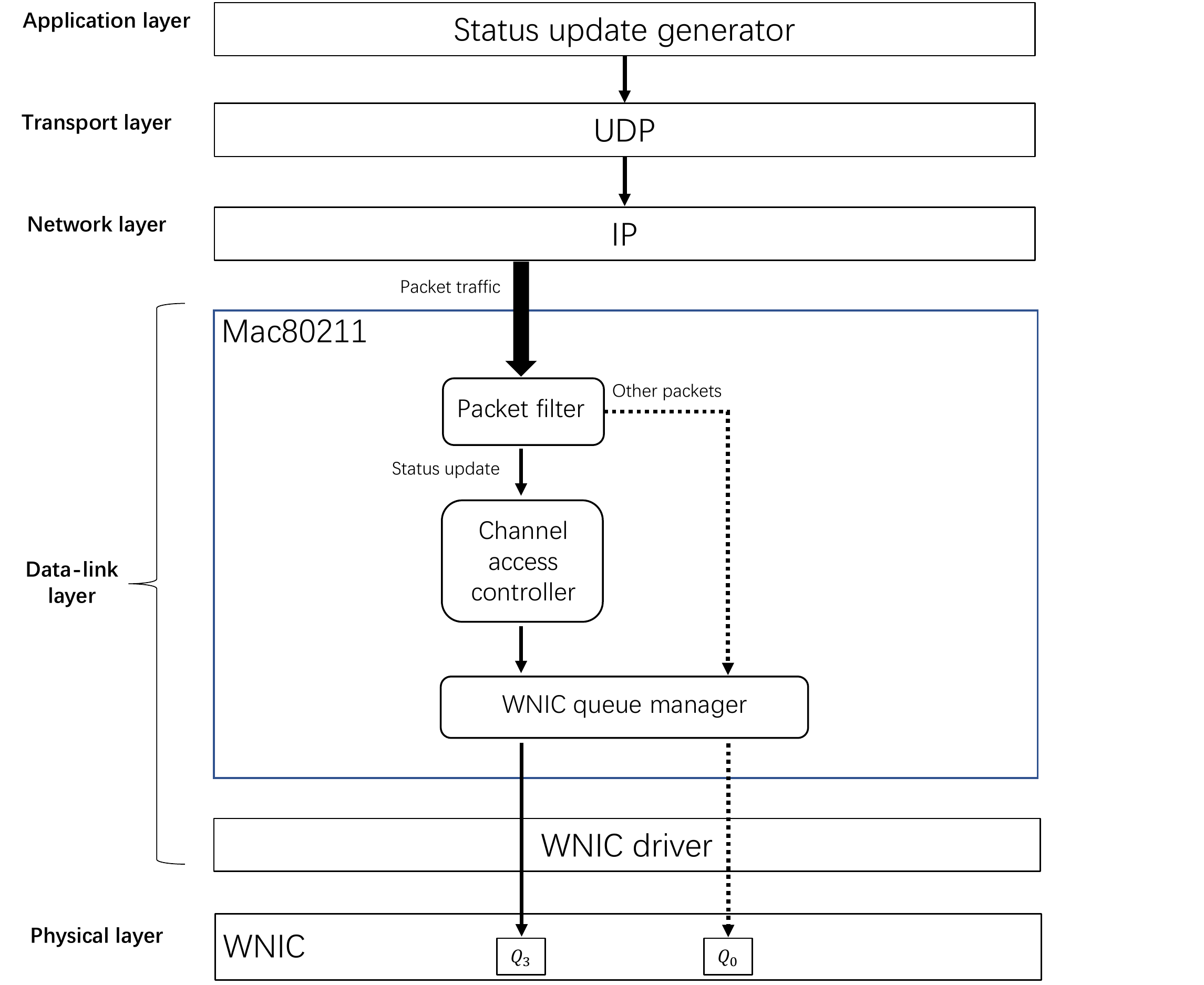}
    \caption{Diagram of Fresh-Fi architecture in implementing the mac80211 subsystem-based LCFS queue over the physical-layer FCFS queues of the WNIC.}
   \label{fig: FRESH-FI_before_TX}
   \vspace{-1em}
\end{figure}

This section discusses the design and implementation details of Fresh-Fi in establishing a mac80211 subsystem-based LCFS queue for timely status updates over the physical-layer FCFS queues of the WNIC, as depicted in Fig. \ref{fig: FRESH-FI_before_TX}. This figure shows three vital modules, and each one of them will be explained along the status update transmission path.

\subsection{Packet Filter}
To prevent accidental injections of non-status update packets into the mac80211 subsystem-based LCFS queue, this module filters out status updates from the incoming traffic of the mac80211 subsystem. It not only ensures that status update packets will not be injected into the same transmission queue as the other packets, but also guarantees that Fresh-Fi will not interfere with the transmissions of packets that require higher reliability. For example, network management protocols, such as address resolution protocol (ARP) and internet control message protocol (ICMP), prioritize packet transmission reliability over speed.

Fresh-Fi implements the packet filter in the \texttt{ieee80211\_subif\_start\_xmit()} function, which serves as the entry point of the mac80211 subsystem. The status update generator at the application layer assigns a unique time-to-live (TTL) value to every generated packet. Based on this TTL value, the packet filter sifts out the status updates from incoming traffic and directs them to the channel access controller, while other packets follow their original transmission paths within the mac80211 subsystem.\footnote{{Alternative filtering strategies can also be utilized. For instance, the packet filter can be configured to identify status update packets by verifying if they are directed towards the predetermined destination IP address and port number.}}

\subsection{Channel Access Controller}
This module is responsible for ensuring the number of status updates stored in the mac80211 subsystem-based LCFS queue does not exceed one, and preventing the WNIC from repeatedly re-transmitting a stale status update.\footnote{{Our current design is optimized for short status update packets. However, we plan to investigate the effects of longer status update packets and the resulting packet segmentation at the transport layer in our future work.}} The channel access controller configures the WNIC to report the transmission status to the mac80211 subsystem once a status update has been delivered. Fresh-Fi uses this message as a feedback packet for the status sampling policy to indicate the transmission completion of the previous status update. Only after receiving a feedback packet, the status sampling policy schedules the next status generation. Additionally, to reduce the wasted time when re-transmitting stale status updates, the channel access controller configures the WNIC to only send each status update once. Before delivering the status update to the WNIC queue manager, the channel access controller backs up this packet for possible re-transmission in the future.

Fresh-Fi implements the channel access controller in the \texttt{ieee80211\_xmit()} function. This module sets the \texttt{IEEE80211\_TX\_CTL\_REQ\_TX\_STATUS} flag in the status update, enabling the WNIC to report the transmission status after finishing transmitting this packet. Moreover, it modifies the source code of the adaptive data rate adjustment algorithm in the \texttt{ieee80211\_get\_tx\_rates()} function, to create a special type of re-transmission table exclusively for status updates. This type of re-transmission table only has a value of one for the maximum re-transmission time of the highest data rate supported by the current channel condition, and zero values for the maximum re-transmission times of the lower data rates. Finally, the channel access controller copies the status update and saves the duplication to its backup buffer, which is an FCFS queue with a size of one, for later usage.

\subsection{WNIC Queue Manager}
Based on the FCFS queue mechanism with multiple priorities implemented in the WNIC, this module injects status updates into the physical-layer FCFS queue with the highest priority, while other packets are injected into the physical-layer FCFS queue with the lowest priority. Since a new status update only comes after the delivery of the previous one, the number of packets stored in the highest priority physical-layer FCFS queue never exceeds one. In addition, due to queue prioritization, the WNIC always transmits the status update earlier than other packets. Consequently, the WNIC queue manager transforms the physical-layer FCFS queue of the WNIC with the highest priority to the mac80211 subsystem-based LCFS queue, which specifically serves for status update transmissions.

Fresh-Fi employs the WNIC queue manager in the \texttt{ieee80211\_tx()} function. The mac80211 subsystem defines the packet priority from zero to fifteen in the packet structure\cite{skb_priority}. By setting the priority field of status updates as fifteen and other packets as zero, the WNIC queue manager ensures that the WNIC injects status updates into the FCFS queue with the highest priority, while other packets are injected into the FCFS queue with the lowest priority. 


\section{Real-time kernel-based cross-layer tunnel in Fresh-Fi}\label{LOCAL FEEDBACK}

\begin{figure}
    \centering
    \includegraphics[width=0.45\textwidth]{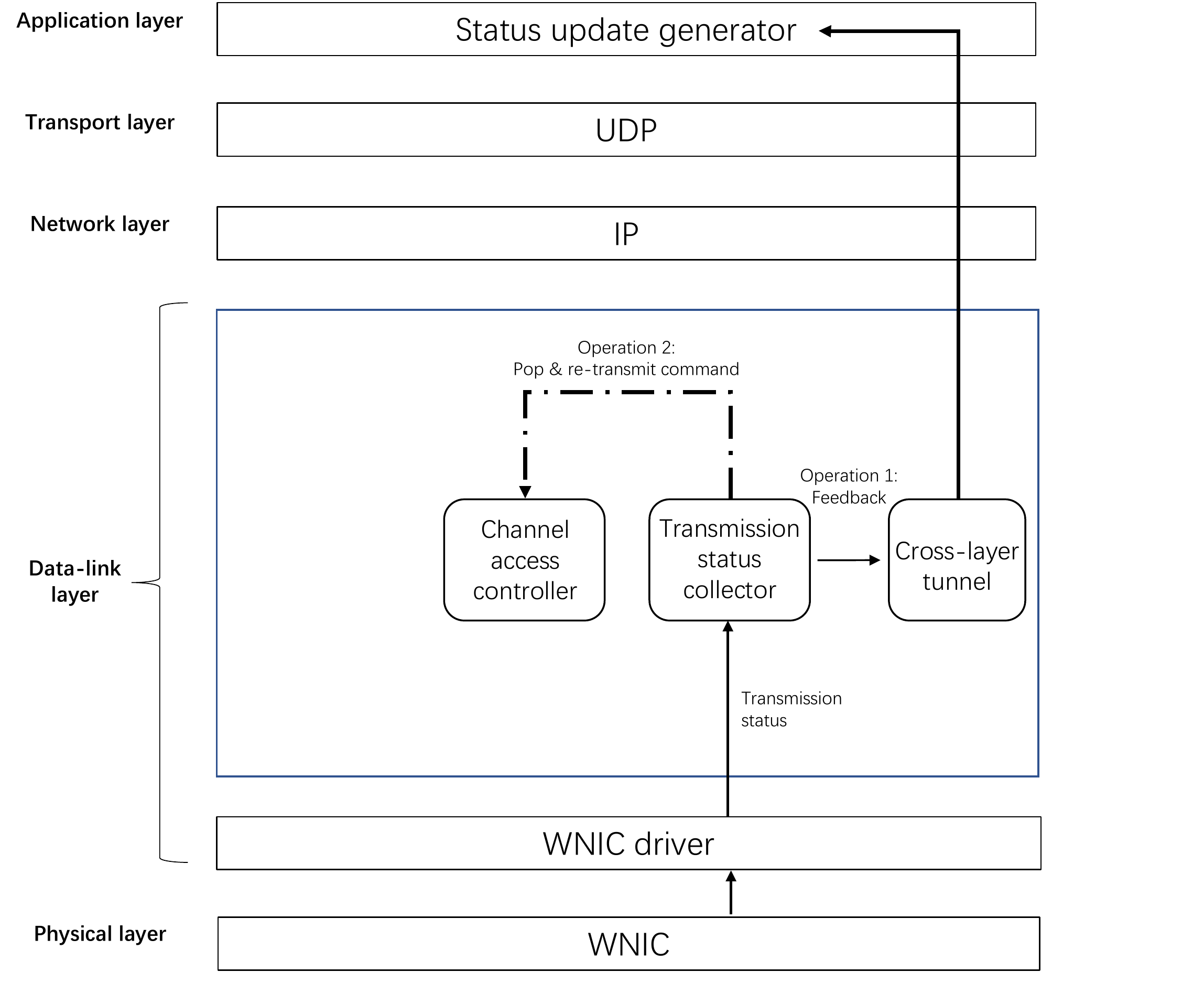}
    \caption{Diagram of Fresh-Fi architecture 
    in implementing the real-time kernel-based cross-layer tunnel.}
   \label{fig: FRESH-FI_after_TX}
   \vspace{-1em}
\end{figure}

To compensate the absence of the transport layer ACKs when using the UDP socket for fast status update transmissions, Fresh-Fi needs to generate feedback packets based on the transmission status reported from the WNIC. These packets are then passed to the status sampling policy through the Fresh-Fi cross-layer tunnel, as the indications of the completion of the previous status update transmission. Fig. \ref{fig: FRESH-FI_after_TX} shows the three key Fresh-Fi modules in building up this cross-layer communication mechanism.

\subsection{Transmission Status Collector}
This module collects the transmission status, then decides whether to generate a feedback packet based on this information or command the channel access controller to re-transmit the previous status update. Recall that the channel access controller sets up the WNIC to send every status update only once, with no guarantee of successful transmission. When a transmission error occurs, it may improve AoI performance by re-transmitting the previous status update. Therefore, the transmission status collector has to decide if a re-transmission is necessary. If the previous status update is successful, then the transmission status collector forwards the transmission status as a feedback packet to the cross-layer tunnel. If unsuccessful, the transmission status collector compares the AoI of the backup status update in the channel access controller to the request-then-arrival (RTA) interval. This interval is defined as the time it takes for a new status update generated in the application layer to reach the mac80211 subsystem after the delivery of a feedback packet by its transmission status collector. If the system time (or local age) of the backup status update is smaller than the RTA interval, then the transmission status collector instructs the channel access controller to resend its stored packet. Otherwise, the transmission status collector sends the transmission status as a feedback packet to the cross-layer tunnel to initiate the generation of a new status update in the application layer.


Fresh-Fi implements the transmission status collector in the \texttt{ieee80211\_tx\_status()} function to collect the transmission status from the WNIC. This module identifies the status update according to the unique TTL value, and then determines whether the transmission of the previous status update is successful or not by checking the \texttt{ACK} flag. The \texttt{ACK} flag is defined in the transmission status, its value of one indicates that the WNIC has received the physical layer ACK, while a value of zero denotes the absence of the physical layer ACK reception. If the \texttt{ACK} equals zero, a transmission error occurs. The transmission status collector calculates the system time of the status update stored in the channel access controller by measuring the time period between the reception of the transmission status and the packet generation time as recorded by the payload. The value of the RTA interval is measured according to the statistical analysis of the status update process for a sufficient long time.


\begin{table}
	\renewcommand{\arraystretch}{1.2}
	\centering
    \small
	\caption{RTA Interval Evaluation Results.}
	\label{tab1}
  \begin{tabular}{c|c|c}
    \toprule
    
     &Without patch&With patch\\
    \hline
    Mean RTA interval&58.17$\mu$s&38.30$\mu$s\\
    \hline
    Standard deviation of RTA interval& 120.40$\mu$s & 71.52$\mu$s\\
    \hline
    Percentage of value $>$100$\mu$s & 6.65\% & 1.33\% 
  \end{tabular}
  \vspace{-1em}
\end{table}

\subsection{Cross-layer Tunnel}
This module is created to transmit feedback packets to the status sampling policy as fast as possible. Fresh-Fi uses Netlink socket technology to establish the cross-layer tunnel, enabling real-time communication between the mac80211 subsystem and the application layer. An effective cross-layer tunnel has to achieve a low RTA interval. We assess the performance of the cross-layer tunnel by conducting a 10-minute point-to-point status update experiment with the just-in-time status sampling policy integrated in the status update generator. The evaluation results are presented in the first column of Table \ref{tab1}. The data shows that the mean RTA interval is 58.17$\mu$s, with a standard deviation of 120.40$\mu$s. The deviation of the recorded data is approximately twice the mean value, indicating the instability in both the cross-layer transmitting and the status update generation process. Furthermore, 6.65\% of the measured RTA interval is larger than 100$\mu$s. To make the RTA interval more stable, we implement the PREEMPT\_RT patch \cite{PREEMPT_RT} in the Linux operating system. The PREEMPT\_RT patch is intended to ensure real-time performance, minimize latency, and provide a rapid response time. We conduct the experiment again for another 10 minutes and record the results in the second column of Table \ref{tab1}. It shows that the installation of PREEMPT\_RT patch decreases the mean RTA interval to 38.30$\mu$s, along with a reduction in the deviation value of approximately 40\% to 71.52$\mu$s. The proportion of RTA interval with a value larger than 100$\mu$s is only 1.33\%. Therefore, we conclude that the implementation of the PREEMPT\_RT patch significantly enhances the overall stability of Fresh-Fi.

\section{Experiment Results}\label{Experiment results}
We begin this section by outlining the design and implementation details of our experimental platform for establishing a single-device\footnote{We note that Fresh-Fi can be directly applied in multi-device WiFi systems. Moreover, the single WiFi device still needs to contend for the channel access with other surrounding WiFi devices.} status update network using COTS WiFi devices. Subsequently, we evaluate and compare the AoI performances of the network with Fresh-Fi and other protocol stacks. Finally, we present the results of a series of experiments aimed at analyzing the effectiveness of various Fresh-Fi modules in improving information freshness.




\subsection{COTS WiFi Device-based Experiment Platform}
We establish our status update system in the dynamic indoor environment by using two MINI PCs running the Ubuntu 18.04 operating system with kernel 4.19.37. The distance between them is around three meters. One MINI PC serves as the AP and the destination node, and the other works as the source node. Both MINI PCs are equipped with the Atheros AR9382 WiFi chip and the supporting ath9k driver, allowing them to communicate in the 2.4 GHz frequency band according to the IEEE802.11g wireless specifications. In the application layer, the source node uses the C language socket library to create a status update generator with the UDP socket for packet transmissions. The status sampling policy is also written in the C language. For Fresh-Fi, we initially integrate the zero-wait sampling policy proposed in \cite{What_is_just_in_time_updating} in the status update generator. This policy schedules the next status generation once it receives a feedback packet from the cross-layer tunnel. Every generated status update is 150 bytes long. It has a unique TTL value\footnote{{We chose to set the TTL value to 53 because it is sufficiently large for transmitting status updates within a network consisting of a single device. How to set the TTL value in larger networks will need further investigation .}} of 53 and its generation timestamp is recorded in the payload. Meanwhile, the destination node runs a status update receiver program written in C language, which extracts the generation timestamps of received status updates to calculate the instantaneous AoI. To calculate the AoI in high precision, we synchronize the source node and destination node using the precision time protocol (PTP) \cite{IEEEPTP}. This synchronization protocol is able to achieve sub-microsecond accuracy and is carried out over the Ethernet connection, without interfering with the status update transmissions over the wireless channel.

\subsection{AoI Performance Comparison between Fresh-Fi and Other baselines}
 We evaluate the AoI performances of three different information freshness-oriented protocol stacks, and display their 10-minute average AoI measured on our experiment platform as shown in Table \ref{tab2}. The considered implementations are:
\begin{itemize}
    \item \textbf{Fresh-Fi}: The source node integrates the zero-wait status sampling policy in the status update generator, and deploys Fresh-Fi as described in Secs. \ref{mac80211 subsystem-based LCFS} and \ref{LOCAL FEEDBACK}.
    \item  \textbf{WiFi UDP}: The source node simply uses the UDP socket for delivering status updates. Once the UDP socket sends the packet to the kernel space, the status update generator immediately produces a new packet for the next transmission.
    \item \textbf{WiFresh APP} \cite{WiFresh}: To our best knowledge, this is the only work that tried to enhance the AoI performance of a COTS WiFi systems. The source node generates status updates at a constant rate R and buffers them in an LCFS queue implemented in the application layer. The destination node sends a polling packet to the source node every 300$\mu$s. Upon receiving a polling packet, the source node sends the latest status update popped from the LCFS queue to the destination node through the UDP socket.

\end{itemize}

\begin{table}
	\renewcommand{\arraystretch}{1.3}
	\caption{Average AoI Comparison among Different Information Freshness-oriented Protocol Stacks}	\centering
	\label{tab2}
  \begin{tabular}{c|c|c|c|c}
    \toprule
     \multicolumn{3}{c|}{WiFresh APP}& \multirow{2}{*}{WiFi UDP}&\multirow{2}{*}{Fresh-Fi}\\
     \cline{1-3}
      {R=5kHz}& {R=6kHz} & {R=7kHz} &\\
     \hline
   22.21ms&25.83ms&28.88ms&35.13ms&1.56ms\\
    \bottomrule
  \end{tabular}
  \vspace{-1em}
\end{table}

Table \ref{tab2} shows that increasing the value of R from 5KHz to 7KHz has little effect on the average AoI of WiFresh APP, which remains around 25ms. In contrast, the average AoI of WiFi UDP is 35.13ms, as shown in the second column. These results suggest that WiFresh APP only reduces the average AoI by approximately 25\%. This is not surprising, as WiFresh APP is implemented solely in the application layer and status updates can still suffer from various delays in the lower layer queues and needs to wait for the polling packet from the destination before pushing down a new status update. By contrast, the third column shows that Fresh-Fi offers a significant improvement in information freshness. Its average AoI equals 1.56ms, which is about one twenty-second of that of the WiFi UDP and one thirteen of that of WiFresh APP. The dramatic AoI improvement of Fresh-Fi is attributed to two reasons. First, Fresh-Fi prohibits the WNIC from repeatedly transmitting the stale status update and thus improves the AoI at the destination node. Second, the mac80211 subsystem-based LCFS queue implemented in Fresh-Fi eliminates the backlogging time suffered by status updates.

\subsection{Effectiveness of Different Fresh-Fi Modules}
Next, we evaluate the effectiveness of different Fresh-Fi modules in decreasing the average AoI. Meanwhile, since the zero-wait status sampling policy is not always optimal as indicated in \cite{zero_wait_not_optimum}, we consider replacing it with the 300$\mu$s-wait status sampling policy. This policy schedules the status update generator to create a new packet 300$\mu$s later, once receiving the feedback packet. The 10-minute average AoI evaluation results are listed in Table \ref{tab3}, which includes the following modified protocol stacks from Fresh-Fi:
\begin{itemize}
    \item \textbf{Fresh-Fi with default WNIC re-transmissions}: The source node integrates the zero-wait status sampling policy in the status update generator, and deploys Fresh-Fi  without modifying the adaptive data rate adjustment algorithm to generate the special type of re-transmission table for status updates.

    \item \textbf{Fresh-Fi without WNIC queue manager}: The source node integrates the zero-wait status sampling policy in the status update generator, and deploys Fresh-Fi without distributing status updates and other packets into different WNIC queues. 

   \item \textbf{Fresh-Fi without cross-layer tunnel}: The source node integrates the zero-wait status sampling policy in the status update generator, and deploys Fresh-Fi without the Netlink sockets for feedback packet transmissions from Fresh-Fi to the status sampling policy. Without feedback on previous status transmission, the status sampling policy can no longer schedule status generations. Thus, the status update generator produces a new packet immediately after the UDP socket sends the previous one into the kernel space.

    \item \textbf{Fresh-Fi with the 300$\mu$s-wait status sampling policy}: Fresh-Fi remains unmodified. The source node integrates the 300$\mu$s-wait status sampling policy, rather than the zero-wait policy, in its status update generator.
\end{itemize}

\begin{table}
	\renewcommand{\arraystretch}{1.3}
	\caption{Average AoI Comparison among the Fresh-Fi Architectures with Different Modifications}	\centering
	\label{tab3}
  \begin{tabular}{c|c}
    \toprule
    Fresh-Fi&1.56ms\\
    \hline
    Fresh-Fi with default WNIC re-transmissions&1.74ms\\
    \hline
    Fresh-Fi without WNIC queue manager&1.62ms\\

    \hline
    Fresh-Fi without cross-layer tunnel&17.27ms\\
    \hline
    Fresh-Fi with the 300$\mu$s-wait status sampling policy&1.14ms\\
    
    \bottomrule
  \end{tabular}
  \vspace{-1em}
\end{table}

Comparing the first and second rows in Table \ref{tab3}, we find that the WNIC re-transmission process leads to around 180$\mu$s increase in the average AoI. This is because the WNIC spends extra time re-transmitting the stale status update whenever the transmission error happens. The comparison between the first and third rows reveals that if status updates lose the transmission priority, the consequent penalty in average AoI is quite negligible, which is around 60$\mu$s. This is because the source node only deploys the status update generator in the application layer, and the network management packet transmissions are sparse in the Linux networking protocol stack. Even without transmission priority, only a few status updates are delayed by the other packets' transmissions. The fourth row shows a dramatic average AoI increase from 1.56ms to 17.27ms. The reason is due to the lack of feedback packets from Fresh-Fi, the status update generator produces a new packet before the previous one is completely delivered. As a result, the increasing backlogging time in the lower layer transmission queues destroys AoI performance.

In Table \ref{tab3}, the last row is the only one providing a smaller average AoI compared to the first row. Its AoI performance improvement is around 25\%, which is due to the receive interruption mitigation mechanism in the WNIC at the destination node. This mechanism defines two thresholds, which are equal to 2000$\mu$s and 500$\mu$s respectively in the ath9k driver, in controlling the interruption mitigation. During high-traffic periods, if the time interval between consecutive received packets is smaller than 500$\mu$s, then the WNIC generates one interruption to report all the packets received in the previous 2000$\mu$s. During low-traffic periods, if the time interval between consecutive received packets is larger than 500$\mu$s, then the WNIC generates one interruption for only one packet received in the previous 500$\mu$s. Since the zero-wait status sampling policy leads to frequent status update receptions with the interval less than 500$\mu$s, these packets are delayed by the receive interruption mitigation mechanism and finally arrive at the receiver program in a burst mode. In contrast, the 300$\mu$s-wait status sampling policy purposely increases the status update reception interval to be large than 500$\mu$s. As a result, the WNIC at the destination node generates interruptions more frequently to report the latest reception. The average AoI decreases from 1.56ms to 1.14ms.

\section{Conclusions and Future Works}\label{Conclusion}
In this paper, we developed Fresh-Fi, the first-of-its-kind information freshness-oriented protocol stack implemented in the Linux mac80211 subsystem that involves careful customization to the lower protocol stack layers of commodity WiFi systems. Fresh-Fi significantly improves the average AoI by designing and implementing two new components: a mac80211 subsystem-based LCFS queue that specifically serves status update transmissions, and a real-time kernel-based cross-layer tunnel that informs the status sampling policy about the previous updating completion for scheduling the next status generation. Our experimental results in a commodity WiFi status update system showed that the implementation of Fresh-Fi improves average AoI performance over 20-fold compared to the baseline simply transmitting status updates through the UDP socket and over 13-fold compared to a state-of-the-art baseline implemented in the application layer. In future work, we plan to refine Fresh-Fi to further improve its AoI performance in multi-user networks. This will involve devising and implementing AoI-aware channel access mechanisms to reduce contention and improve overall network efficiency. Additionally, we aim to enhance Fresh-Fi by providing more physical layer information, such as packet transmission time and transmission rate, to the status sampler, so that the waiting time between status generations can be dynamically adjusted to further optimize AoI performance.



\ifCLASSOPTIONcaptionsoff
  \newpage
\fi

\bibliographystyle{IEEEtran}
\bibliography{References}

\end{document}